\documentclass[final,3p,times,twocolumn]{elsarticle}

\usepackage{graphicx}
\usepackage{amssymb,amsmath}
\usepackage{color}

\usepackage{lineno}

\begin{document}

\begin{frontmatter}

\title{Absence of a day--night asymmetry in the $^7$Be solar neutrino rate in Borexino}

\author[mil]{G. Bellini}
\author[pu]{J. Benziger}
\author[ham]{D. Bick}
\author[mil]{S. Bonetti}
\author[lngs]{G. Bonfini}
\author[mil]{M. Buizza Avanzini}
\author[mil]{B. Caccianiga}
\author[um]{L.~Cadonati}
\author[pu]{F.~Calaprice}
\author[ge]{C.~Carraro}
\author[mil]{P.~Cavalcante}
\author[pu]{A.~Chavarria}
\author[mil]{D. D{\textquoteright}Angelo}
\author[ge]{S.~Davini}
\author[stp]{A.~Derbin}
\author[kur]{A.~Etenko}
\author[tum]{F.~von~Feilitzsch}
\author[dub]{K.~Fomenko}
\author[apc]{D.~Franco}
\author[pu]{C. Galbiati}
\author[lngs]{S. Gazzana}
\author[lngs]{C.~Ghiano}
\author[mil]{M.~Giammarchi}
\author[tum]{M.~G\"{o}ger-Neff}
\author[pu,lngs]{A.~Goretti}
\author[pu]{L.~Grandi}
\author[ge]{E. Guardincerri}
\author[vt]{S. Hardy}
\author[lngs]{Aldo Ianni}
\author[pu]{Andrea Ianni}
\author[kiev]{V.~Kobychev}
\author[dub]{D.~Korablev}
\author[lngs]{G.~Korga}
\author[lngs]{Y. Koshio}
\author[apc]{D. Kryn}
\author[lngs]{M. Laubenstein}
\author[tum]{T. Lewke}
\author[kur]{E. Litvinovich}
\author[pu]{B.~Loer}
\author[mil]{P.~Lombardi}
\author[lngs]{F.~Lombardi}
\author[mil]{L. Ludhova}
\author[kur]{I. Machulin}
\author[vt]{S. Manecki}
\author[mpik]{W. Maneschg}
\author[ge]{G.~Manuzio}
\author[tum]{Q.~Meindl}
\author[mi]{E.~Meroni}
\author[mi]{L.~Miramonti}
\author[jag]{M. Misiaszek}
\author[lngs,pu]{D. Montanari}
\author[pu]{P. Mosteiro}
\author[stp]{V. Muratova}
\author[tum]{L.~Oberauer}
\author[apc]{M.~Obolensky}
\author[per]{F.~Ortica}
\author[ge]{M.~Pallavicini}
\author[vt]{L.~Papp}
\author[carlos]{C. Pe\~na-Garay} 
\author[mil]{L.~Perasso}
\author[ge]{S.~Perasso}
\author[um]{A.~Pocar}
\author[vt]{R.S.~Raghavan}
\author[mil]{G.~Ranucci}
\author[lngs]{A.~Razeto}
\author[mil]{A.~Re}
\author[per]{A. Romani}
\author[kur]{A. Sabelnikov}
\author[pu]{R. Saldanha}
\author[ge]{C. Salvo}
\author[mpik]{S.~Sch\"onert}
\author[mpik]{H.~Simgen}
\author[kur]{M.~Skorokhvatov}
\author[dub]{O.~Smirnov}
\author[dub]{A. Sotnikov}
\author[kur]{S. Sukhotin}
\author[lngs,kur]{Y. Suvorov}
\author[lngs]{R.~Tartaglia}
\author[ge]{G.~Testera}
\author[apc]{D.~Vignaud}
\author[vt]{R.B.~Vogelaar}
\author[tum]{J.~Winter}
\author[jag]{M. Wojcik}
\author[pu]{A. Wright}
\author[tum]{M. Wurm}
\author[pu]{J.~Xu}
\author[dub]{O.~Zaimidoroga}
\author[ge]{S.~Zavatarelli}
\author[jag]{G.~Zuzel}

\address[apc]{APC, Laboratoire AstroParticule et Cosmologie, 75231 Paris cedex 13, France}
\address[dub]{Joint Institute for Nuclear Research, Dubna 141980, Russia}
\address[ge]{Dipartimento di Fisica, Universit\'{a} e INFN, Genova 16146, Italy}
\address[jag]{M. Smoluchowski Institute of Physics, Jagiellonian University, Krakow, 30059, Poland}
\address[lngs]{INFN Laboratori Nazionali del Gran Sasso, Assergi 67010, Italy}
\address[kiev]{Kiev Institute for Nuclear Research, Kiev 06380, Ukraine}
\address[kur]{NRC Kurchatov Institute, Moscow 123182, Russia}
\address[mil]{Dipartimento di Fisica, Universit\'{a} degli Studi e INFN, Milano 20133, Italy}
\address[mpik]{Max-Plank-Institut f\"{u}r Kernphysik, Heidelberg 69029, Germany }
\address[per]{Dipartimento di Chimica, Universit\'{a} e INFN, Perugia 06123, Italy }
\address[puchem]{Chemical Engineering Department, Princeton University, Princeton, NJ 08544, USA }
\address[pu]{Physics Department, Princeton University, Princeton, NJ 08544, USA }
\address[stp]{St. Petersburg Nuclear Physics Institute, Gatchina 188350, Russia }
\address[tum]{Physik Department, Technische Universit\"{a}t M\"{u}nchen, Garching 85747, Germany }
\address[carlos]{Instituto de F\`isica Corpuscular, CSIC-UVEG, Valencia E-46071, Espa\~na }
\address[um]{Physics Department, University of Massachusetts, Amherst 01003, USA }
\address[vt]{Physics Department, Virginia Polytechnic Institute and State University, Blacksburg, VA 24061, USA }
\address[ham]{ Institut f\"ur Experimentalphysik, Universit\"at Hamburg, Germany }

\begin{abstract}
We report the result of a search for a day--night asymmetry in the 
$^7$Be solar neutrino interaction rate in the Borexino detector at 
the Laboratori Nazionali del Gran Sasso (LNGS) in Italy. 
The measured asymmetry is A$_{dn}$~=~0.001 $\pm$ 0.012 (stat) $\pm$ 0.007 (syst), 
in agreement with the prediction of MSW-LMA solution for neutrino
oscillations. This result disfavors MSW oscillations with mixing
parameters in the LOW region at more than 8.5$\,\sigma$.
This region is, for the first time, strongly disfavored without
the use of reactor anti-neutrino data and therefore the assumption of CPT symmetry. 
The result can also be used to constrain some neutrino oscillation scenarios involving new physics.

\end{abstract}


\end{frontmatter}

\linenumbers

In the last two decades solar
neutrino~\cite{bib:RadioChemical,bib:SK,bib:SNO} and reactor
anti-neutrino \cite{bib:Kamland} experiments have demonstrated that solar
electron neutrinos undergo flavor conversion along their trip  from
the Sun's core to the Earth. The conversion is well described by the
so-called Mikheyev-Smirnov-Wolfenstein (MSW) matter-enhanced neutrino
oscillations \cite{bib:MSW} with Large Mixing Angle (LMA)
oscillation parameters. A generic feature of matter-enhanced neutrino
oscillations is the potential for the coherent re-generation of
the $\nu_e$ flavor eigenstate when solar neutrinos propagate through
the Earth \cite{bib:Regener}, as they do during the night. 
Thus, there is the potential for those
solar neutrino experiments that are principally, or entirely,
sensitive to $\nu_e$ to detect different solar
neutrino interaction rates during the day and during the night. Solar neutrino
day-night asymmetry measurements are sensitive to both $\nu_e$ appearance and disappearance.

The magnitude of this day-night effect is expected to
depend on both neutrino energy and the neutrino oscillation
parameters. Previous experiments \cite{bib:SNOdn, bib:SKdn} have shown
that for high energy ($\sim$5-15 MeV) solar neutrinos, the day--night asymmetry is less than a 
few percent, in agreement with the MSW-LMA
prediction. 
At lower neutrino energies (around 1 MeV), the
predicted day--night asymmetry for MSW-LMA is also small ($<$0.1\%) \cite{bib:day-night-LMA};
however, other scenarios including different MSW solutions and neutrino
mixing involving new physics \cite{bib:mavan} predict much larger day--night effects. For example, in the so-called 
LOW region (10$^{-8}$ eV$^2$$<$$\delta m^2$$<$10$^{-6}$ eV$^2$)
of MSW parameter space, which is currently strongly disfavored only
by the KamLAND anti-neutrino measurement under the assumption of CPT symmetry,
the day-night asymmetry would range between about 10\% and 80\% for
neutrino energies near 1 MeV.  
We present here
the first measurements sensitive to the day--night asymmetry for solar neutrinos
below 1 MeV. This result is an essentially new and independent way
to probe the MSW-LMA prediction and is
potentially sensitive to new physics affecting the propagation of low energy electron neutrino in matter.
Particularly, this result is independent from the KamLAND
measurement, which probes anti-neutrino interactions at higher energies ($>$1.8 MeV).

The Borexino experiment at LNGS detects low energy solar neutrinos by
means of their elastic scattering on electrons in a large volume
liquid scintillator detector. Real-time detection (with $\approx$1
$\mu$s absolute time resolution) of all events is made by collecting
the scintillation light with a large set of photomultipliers. The
very low intrinsic radioactivity of the scintillator and of the materials
surrounding it allows a clean spectral separation
between the neutrino signals and the residual
background. As the neutrino-electron elastic scattering cross section
is different for $\nu_e$ and $\nu_\mu$-$\nu_\tau$, Borexino can measure
the electron neutrino survival probability and is, as a result,
sensitive to the day-night effect. 

We recently released a precise measurement of the $^7$Be neutrino
interaction rate in Borexino with a total uncertainty less than
5\%~\cite{bib:BxBe3}. In this Letter, we present a study of the
day-night asymmetry in the same $^7$Be solar neutrino rate, placing a 
stringent limit on the size of the possible effect. We show that this 
limit improves the constraint on the solar neutrino oscillation parameters 
from solar neutrino experiments alone and excludes new physics 
scenarios that cannot be rejected with the currently available data.

\begin{figure}[t]
\begin{center}
\includegraphics[width=0.52\textwidth]{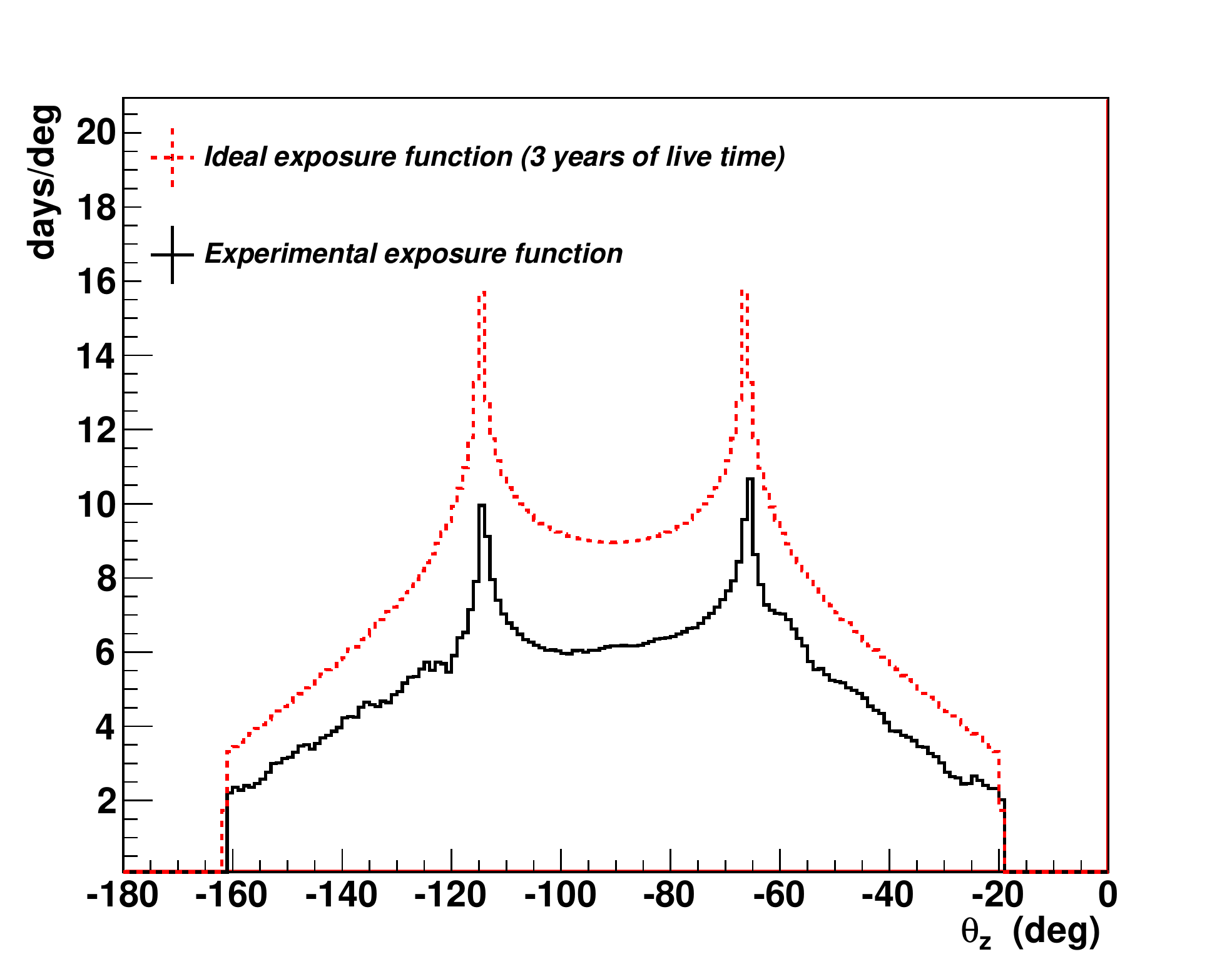}
\caption{The experimental exposure function (black continuous line) and the ideal exposure function (red dotted line).
The interval from $-180^\circ$ to $-90^\circ$  corresponds to day time and the one from   $-90^\circ$ to $0^\circ$ to night time. We recall that at LNGS latitude the Sun is never at the zenith.}
\label{fig:zenith}
\end{center}
\end{figure}

The Borexino detector \cite{bib:nim1,bib:nim2} is located in Hall C of the Laboratori Nazionali del Gran Sasso (latitude 42.4275$^\circ$ N) in Italy and has taken data since May 2007.
The sensitive detector consists of $\sim$278~tons of very pure organic liquid scintillator contained in a 4.25 m radius nylon vessel. The scintillator is viewed by 2212 photomultipliers and is shielded against external neutrons and $\gamma$ radiation~\cite{bib:muon-paper}. The energy of each candidate event is measured by the total amount of collected light, 
while the position of the event is reconstructed using the time-of-flight of the light to the photomultipliers.

The data used in this analysis were collected between May 16$^{th}$,
2007 and May 8$^{th}$, 2010 and correspond to 740.88 live days after
applying the data selection cuts.
We define ``day'' and ``night'' using $\theta_z$,
the angle between the vertical $z$-axis of the detector (positive upward) and
the vector pointing to the detector from the Sun, following \cite{bib:SK}. Note that, with this 
definition, cos\,$\theta_z$ is negative during the day and positive
during the night. The distance that the neutrinos propagate within the
Earth is small for negative
cos\,$\theta_z$ (the $\sim$1.4~km LNGS overburden) and ranges up to 12049 km
for positive cos\,$\theta_z$. Our day and night
livetimes were 360.25 and 380.63 days, respectively.
The distribution of $\theta_z$ corresponding to the live time
(experimental exposure function) is shown in Fig.~\ref{fig:zenith}
and its asymmetry with respect to -90$^\circ$ is mainly due to maintenance and calibration activities which are normally carried out during the day. 

As discussed in~\cite{bib:BxBe3}, scintillation events due to $^7$Be
solar neutrinos cannot be distinguished from background
events (cosmogenics and radioactivity) on an event-by-event basis.
The signal and background contributions are therefore determined using
a spectral fit
to the energy spectrum of the events reconstructed within a suitable
fiducial volume (86.01~m$^3$ in \cite{bib:BxBe3}), and passing a series of cuts which eliminate muons
and short--lived cosmogenic events, time correlated background events, and spurious
noise events (details of this event selection will be published in
\cite{bib:Be7Long}). The experimental signature of the mono--energetic
862\,keV $^7$Be solar neutrinos is a Compton-like electron scattering
``shoulder''  at approximately 660\,keV. 
%

%
In the analysis reported here we use a spherical fiducial volume significantly larger than the one
used in ~\cite{bib:BxBe3} in order to increase the size of the data sample. This choice
is justified by the fact that the additional external background that
enters this larger fiducial volume is due to gamma radioactivity
emitted by the materials surrounding the scintillator volume. As this
background is expected to be the same during day and night, it should
not affect the day--night asymmetry\footnote{As explained later on, not all
backgrounds relevant for this analysis are the same during day and night. Particularly,
the background induced by $^{210}$Po $\alpha$s is not the same
because of the long $^{210}$Po lifetime and of the different length of days and nights in
summer or winter.}.

We determined that our
sensitivity to the day--night effect is maximized by a 3.3\,m fiducial
radius, which gives a 132.5\,ton fiducial mass containing
4.978$\times$10$^{31}$~e$^{-}$. With this choice of fiducial mass,
the signal--to--background (S/B) ratio
in the ``$^7$Be neutrino energy window'' (550 to 800 keV) is
0.70~$\pm$~0.04. This value is smaller than the one in
\cite{bib:BxBe3} due to the increase in spatially non-uniform
backgrounds produced by external gamma rays and $^{222}$Rn events.

The day--night asymmetry, A$_{dn}$, of the $^7$Be count rate is defined as:
\begin{equation}
A_{dn} = 2~\frac{R_N - R_D}{R_N + R_D} = \frac{R_{\mathrm{diff}}}{\left<R\right>}
\label{eq1}
\end{equation}
where $R_N$ and $R_D$ are the $^7$Be neutrino interaction rates during
the night and the day, respectively, $R_{\mathrm{diff}}$ is their difference, and $\left<R\right>$ is their mean. 

\begin{figure}[h]
\begin{center}
\includegraphics[width=0.53\textwidth]{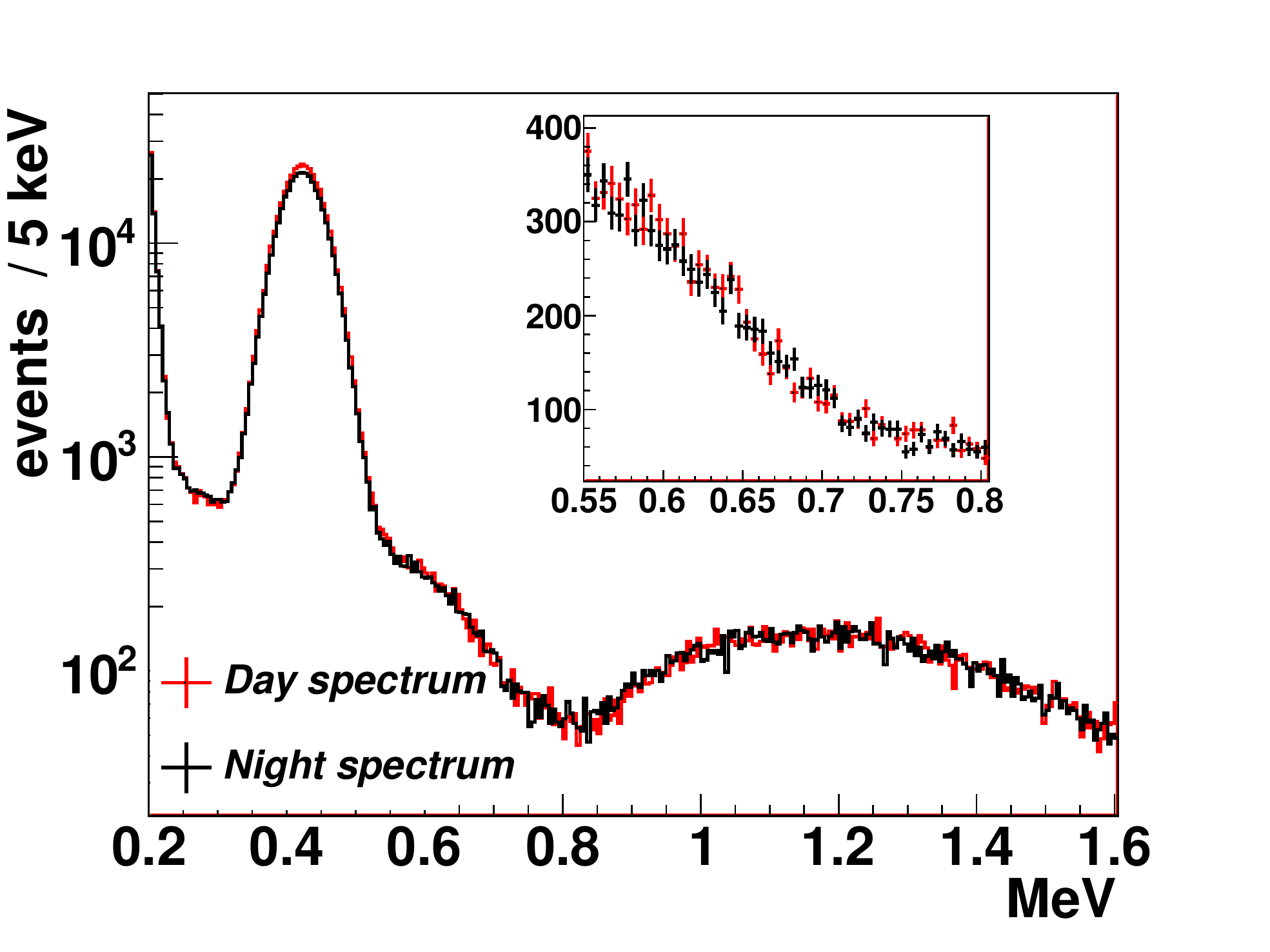}
\caption{The energy spectrum of events during day (red) and night (black) normalized to the day live--time in 
the enlarged FV. The insert shows the $^7$Be neutrino energy window. See \cite{bib:BxBe3} for details on this spectral shape.}
\label{fig:spectra-DN-1}
\end{center}
\end{figure}
\begin{figure}[t]
\begin{center}
\includegraphics[width=0.52\textwidth]{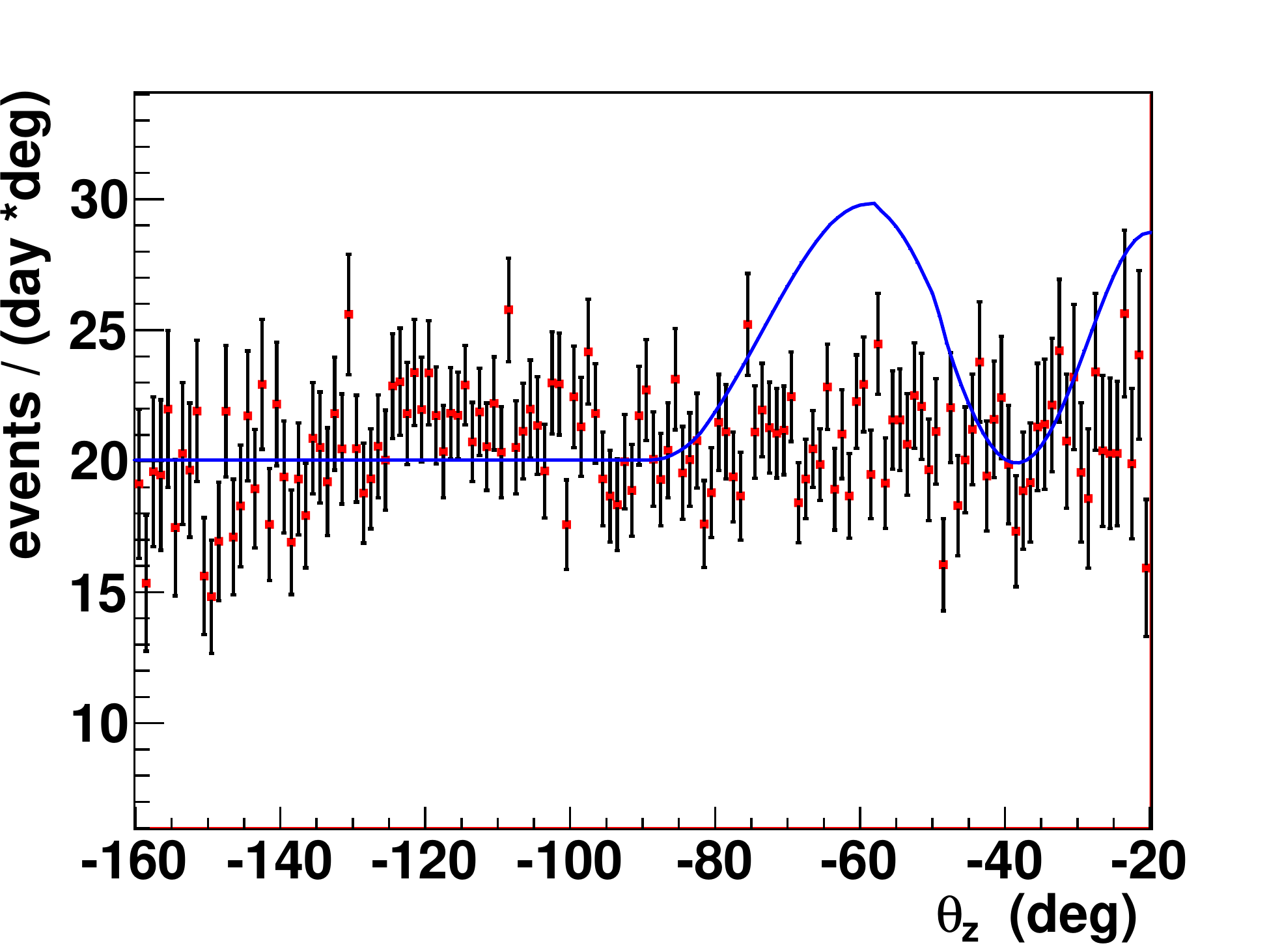}
\end{center}
\caption{Normalized $\theta_z$--angle distribution of the events in the
FV in the $^7$Be neutrino energy window. The effect of the Earth's elliptical orbit has been removed. The blue line is the expected effect with the LOW solution ($\Delta$m$^2_{12}$=1.0 $\cdot$ 10$^{-7}$ eV$^2$  and tan$^2(\theta_{12})$ = 0.955).}
\label{fig:zenithNeutrino}
\end{figure}

Fig. \ref{fig:spectra-DN-1} shows the day and night energy spectra superimposed
and normalized to the same live--time (the day one), while
Fig.~\ref{fig:zenithNeutrino} shows the $\theta_z$
distribution of the events in the $^7$Be neutrino energy window normalized by the experimental exposure
function. By using the total $^7$Be count rate measured in \cite{bib:BxBe3}, a correction has been applied to the exposure function to account for the annual modulation of the neutrino flux due to the seasonal 
variation of the Earth-Sun distance. 
Before correction, the asymmetric distribution of our day and night livetime
throughout the year is expected to increase the measured $^7$Be neutrino count rate by
0.37\% during the night and decrease it by 0.39\% during the day.
The day and night spectra in Fig~ \ref{fig:spectra-DN-1} are
statistically identical, as proved by the fit to the data shown in Fig.~\ref{fig:zenithNeutrino}. Indeed, 
by fitting with a constant distribution the data in Fig.~\ref{fig:zenithNeutrino} we obtain a $\chi^2$ probability = 0.44. Any deviation from a straight line would be a signature of day--night modulation. For illustration, we include in
Fig.~\ref{fig:zenithNeutrino} the expected shape for the LOW solution
($\Delta$m$^2_{12}$=1.0 $\cdot$ 10$^{-7}$ eV$^2$  and tan$^2(\theta_{12})$ = 0.955).
Fitting the distribution with a flat straight line yields $\chi^2$/ndf = 141.1/139, 
showing that the data are consistent with the no day--night effect hypothesis. 

One way to quantitatively constrain A$_{dn}$ is to
determine $R_D$ and $R_N$ separately by independently fitting
the day and night spectra using the same spectral fitting technique used
in determining the total $^7$Be flux
in~\cite{bib:BxBe3} and then comparing the results
using Eq. 1. Note that because these neutrinos are mono--energetic,
we expect the shape of the $^7$Be electron recoil spectrum to be
identical during day and night. This yields $A_{dn}=0.007 \pm 0.073$.
This method has the virtue of allowing for the possibility of different background rates during
day and night. However, this analysis is less sensitive than the one described
below and is not used for the final result. 

A stronger constraint on A$_{dn}$ is obtained by making the very reasonable assumption that the main backgrounds that limit the sensitivity in \cite{bib:BxBe3} ($^{85}$Kr and $^{210}$Bi) are the same during day and night. With this assumption, A$_{dn}$ is obtained by subtracting
the day and night spectra (normalized to the day live time) following the
second term in Eq.~(\ref{eq1}) and then searching for a residual
component having the shape of the electron recoil spectrum due to
$^7$Be neutrinos.
If A$_{dn}$=0 and the background count rates were constant in time
the subtracted spectrum would be flat.

\begin{figure}[t]
\begin{center}
\includegraphics[width=0.53\textwidth]{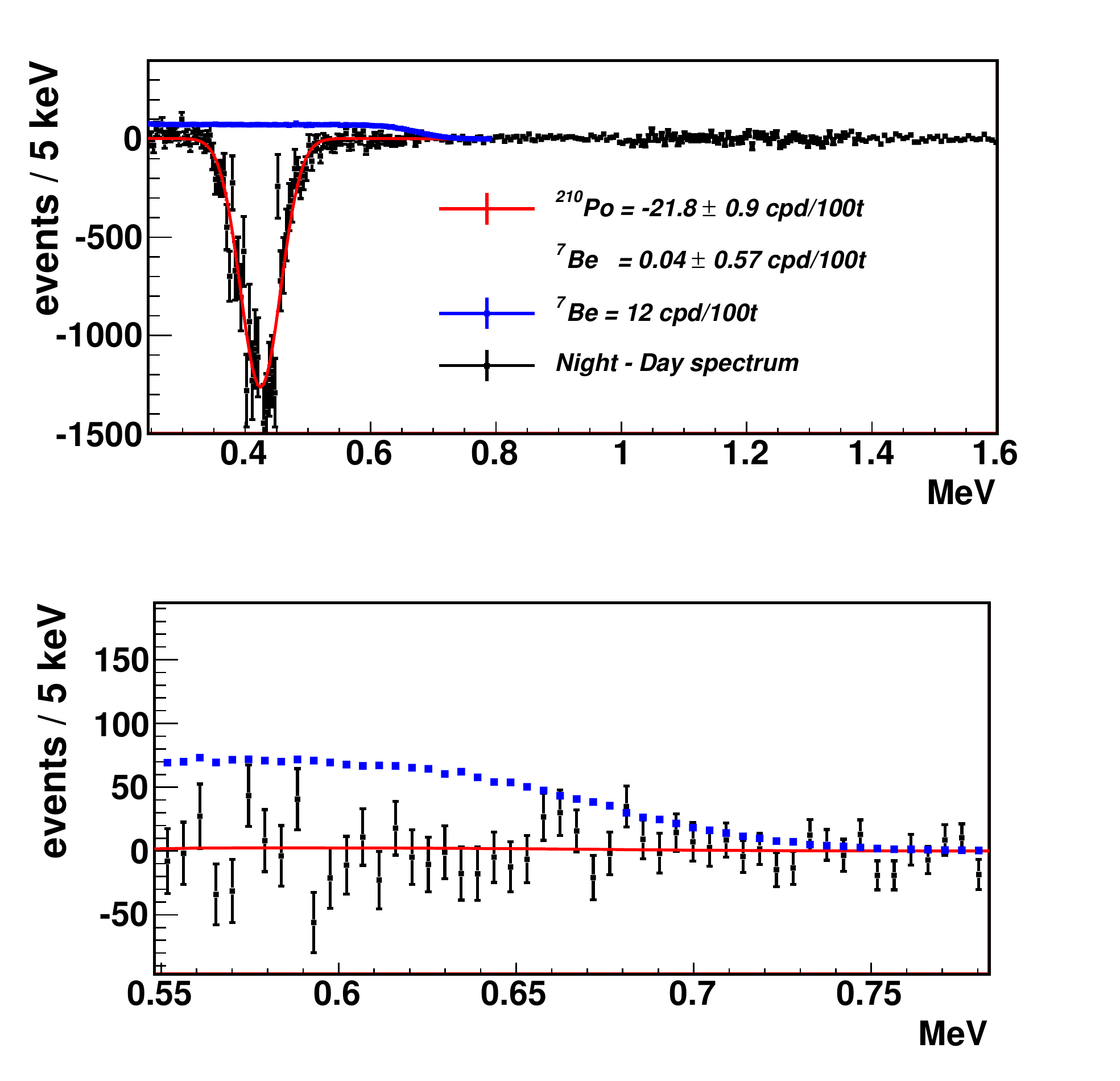}
\caption{Difference of night and day spectra in the
FV. The fit is performed in the energy region between 0.25 and 0.8 MeV with the residual $^{210}$Po spectrum and the electron recoil  spectrum due to the $^7$Be solar neutrino interaction. The fit results
are in cpd/100 t. The top panel shows an extended energy range including the region dominated by the $^{11}$C background while the bottom panel is a zoom of the $^7$Be energy window between 0.55 and 0.8 MeV. The blue curve shows the shape of electron recoil spectrum that would be seen assuming the LOW solution as in Fig. \ref{fig:zenithNeutrino}. }
\label{fig:fit-day-night}
\end{center}
\end{figure}

The subtracted spectrum is shown in  Fig.~\ref{fig:fit-day-night}, where the lower plot is a zoom of the 
upper one in the energy region between 0.55 and 0.8 MeV.
The result is a flat spectrum, consistent with zero, except for a clear negative $^{210}$Po peak visible in the low energy region.
This negative peak arises because the $^{210}$Po
background count rate in Borexino is decaying in time ($\tau_{1/2}$ =
138.38 days), and the day and night livetime are not evenly
distributed over the 3 years of data taking. The $^{210}$Po
count rate was highest at the time of the initial filling in May 2007, and has since
decayed. Therefore, the $^{210}$Po count rate has been higher on average during the
summers (when days are longer), leading to a noticeable effect in the
subtracted spectrum. 
This effect is taken into account by including both the $^{210}$Po and
$^7$Be spectral shapes in the fit. Fitting between 0.25 and 0.8 MeV, we obtain $R_{\mathrm{diff}}$ =
0.04$\pm$0.57 (stat) cpd/100 t. The amplitude of the resulting electron recoil spectrum induced by the interaction of $^{7}$Be neutrinos is too small to be shown in Fig.~\ref{fig:fit-day-night}. In order to see its spectral shape  
we plot the recoil  spectrum with an amplitude corresponding to the expected day--night asymmetry for the LOW solution.

The $R_{\mathrm{diff}}$ result is confirmed by removing alpha events from the day and
night spectra using a pulse
shape analysis based statistical subtraction
technique~\cite{bib:BxBe3} before creating the difference
spectrum. In this case, no residual $^{210}$Po peak is expected or
observed in the difference spectrum. Fitting the data between
0.25 and 0.8 MeV using only the $^7$Be recoil shape yields
a result  consistent with the previous one. 
The difference in the central values is included in the systematic uncertainty.

Using $\left<R\right>$ = 46 $\pm$ 1.5 (stat) $^{+1.6}_{-1.5}$ (syst) ~cpd/100
t~\cite{bib:BxBe3} we obtain A$_{dn}$ = 0.001 $\pm$ 0.012
(stat) $\pm$ 0.007 (syst) from Eq. 1. The statistical error in
$A_{dn}$ is given by
$$\sigma_{A_{dn}} = \frac{R_{diff}}{\left<R\right>} \sqrt{\left(\frac{\sigma_{diff}^2}{R_{diff}^2} + \frac {\sigma^2(<R>)}{<R>^2} \right)} \simeq \frac {\sigma(R_{diff})}{\left<R\right>} $$
because the total relative experimental error associated with $\left<R\right>$ is
negligible with respect to $\frac {\sigma (R_{diff})}{R_{diff}}$. 

The main systematic errors are listed in Table
\ref{table:sys-error}. The dominant uncertainties are associated with
the difference between the $R_{\mathrm{diff}}$ central values obtained
with and without statistical subtraction of the $\alpha$ events, and
the maximum effect on $R_{\mathrm{diff}}$ from potential small changes
in the $^{210}$Bi background in the detector. These uncertainties will
be detailed in \cite{bib:Be7Long}.

\begin{table}[b]
\begin{center}
\begin{tabular}{|c|c|} \hline
Source of error     &    Error on $A_{dn}$  \\ \hline
Live--time  &  $<\,$5$\cdot$10$^{-4}$ \\
Cut efficiencies  & 0.001 \\
Variation of $^{210}$Bi with time & $\pm$0.005 \\ 
Fit procedure & $\pm$ 0.005 \\ \hline
Total systematic error & 0.007 \\ \hline
\end{tabular}
\end{center}
\caption{List of systematic errors on $A_{dn}$.}
\label{table:sys-error}
\end{table}

\begin{figure}[t]
\begin{center}
\includegraphics[width=0.48\textwidth]{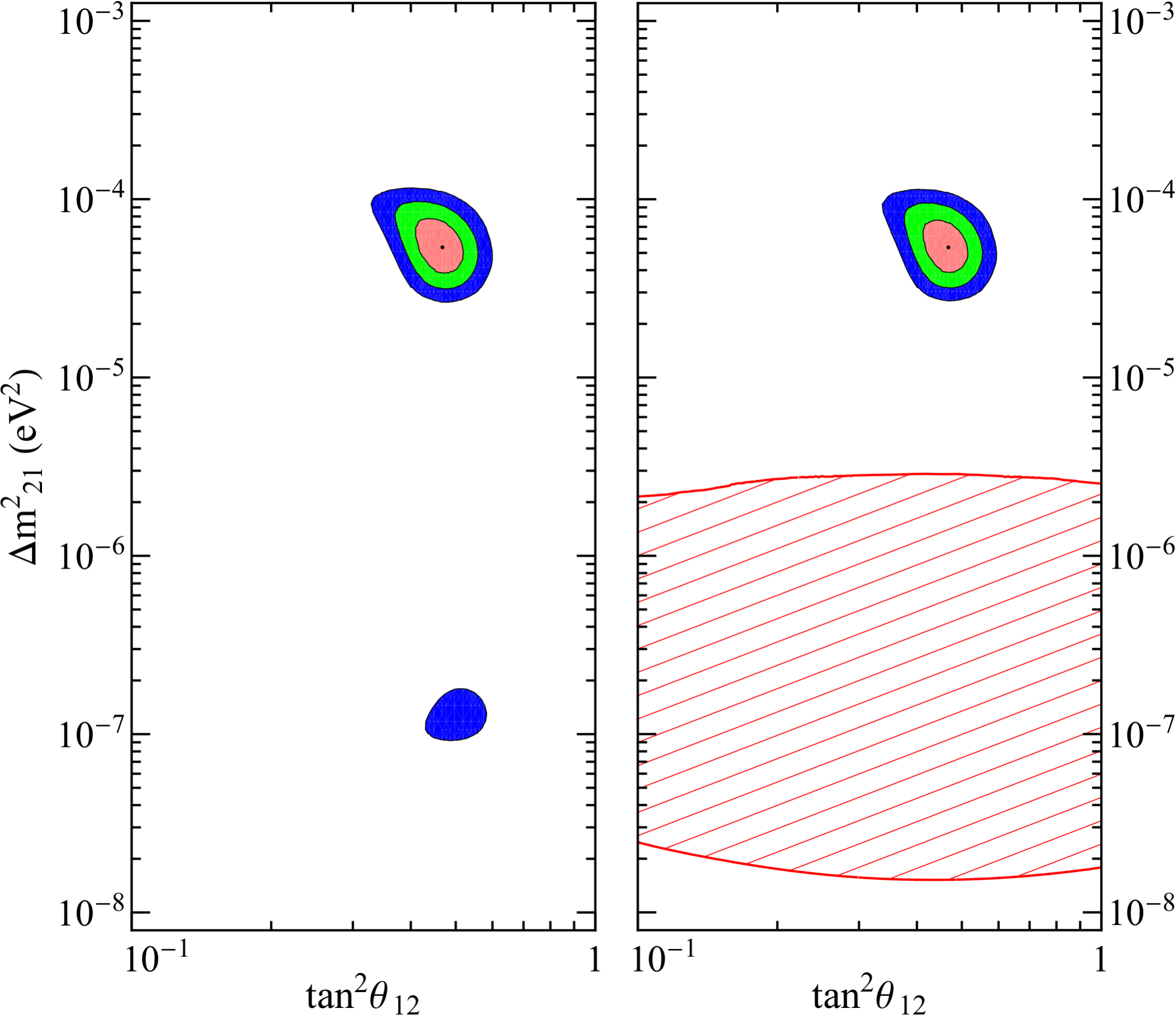}
\caption{
Neutrino oscillations parameter estimation in three solar neutrino data analyses (with 2 d.o.f.): 
1) 99.73\% c.l. excluded region by the Borexino $^7$Be day--night data (hatched red region in the right panel); 
2) 68.27\%, 95.45\%, and 99.73\% c.l.  allowed regions by the solar neutrino data without Borexino data 
(left panel); 
3) Same c.l. allowed regions by all solar neutrino data including Borexino (filled contours in right panel). 
The best fit point in the left (right) panel is $\Delta$m$^2$ = (5.2 $^{+1.6}_{-0.9}$) $\cdot 10^{-5}$, tan$^2 \theta$=0.47 
$^{+0.04}_{-0.03}$ (0.46 $^{+0.04}_{-0.03}$). 
The LOW region is strongly excluded by the $^7$Be day--night data while the allowed LMA parameter region does 
not change significantly with the inclusion of the new data. }
\label{fig:global}
\end{center}
\end{figure}

This new tight constraint on the day-night effect in $^7$Be solar
neutrinos has interesting implications on our understanding of
neutrino oscillations. To investigate this, we calculated
the expected day--night asymmetry for 862\,keV neutrinos under
different combinations of mixing parameters in the MSW
oscillation scenario. 
The comparison of these predictions with our
experimental number is displayed on the right panel of Fig.~\ref{fig:global}. The red
region is excluded at 99.73\% c.l. (2 d.o.f.).
In particular, the minimum day--night
asymmetry expected in the LOW region (10$^{-8}$ eV$^2$
$\lessapprox \Delta$m$^2 \lessapprox 10^{-6}$ eV$^2$) is 0.117,
which is more than 8.5$\,\sigma$ away from our measurement, assuming
gaussian errors for $A_{dn}$.  

This effect can also be seen in a global analysis of all solar neutrino
data. We have carried out such an analysis, assuming two
neutrino oscillations (i.e. $\theta_{13}=0$, we have checked that the
inclusion of the third family does not change any of the conclusions
and will be published in \cite{bib:Be7Long}), including the
radiochemical data \cite{bib:RadioChemical}, the  Super-Kamiokande
phase I and phase III data \cite{bib:SK}, and the SNO LETA data and phase III rates \cite{bib:SNO}. The analysis takes into account the experimental errors (the
systematic and statistical errors summed in quadrature) and the theoretical errors in the total count rates, including the
correlation of the $^7$Be and $^8$B theoretical fluxes
~\cite{bib:solar-model}. We use flux predictions from a recent high metallicity standard solar
model \cite{bib:sun-metallicity} and we include the bin-to-bin
correlations in the uncertainties in the predicted $^8$B neutrino recoil
spectrum resulting from the  uncertainties in the predicted neutrino
spectrum, and from energy threshold uncertainties and energy resolution in the experiments. 

The left panel of Fig.~\ref{fig:global} shows the 68.27, 95.45 and
99.73\% c.l. neutrino mixing parameter regions allowed by all solar neutrino
data without Borexino. The best-fit point is in the LMA region
($\Delta$m$^2$ =  (5.2 $^{+1.6}_{-0.9}$) $\cdot$10$^{-5}$ eV$^2$ and $\tan^2\theta$ =0.47$^{+0.04}_{-0.03}$) 
and a small portion of the LOW region is still allowed at $\Delta \chi^2 =11.83$. 

The right panel of Fig.~\ref{fig:global} shows the regions of allowed
parameter space after adding the Borexino
data (the $^7$Be total count rate ~\cite{bib:BxBe3}, the day--night
asymmetry reported in this paper, and the $^8$B total count rate above 3 MeV (0.22 $\pm$ 0.04 (stat) $\pm$ 0.01 (syst)) cpd/100 t and spectral shape 
(5 bins from 3 to 13 MeV)~\cite{bib:BxB8}) to the analysis.  The LMA
region is only slightly modified (the new best fit point is
$\Delta$m$^2$~=~(5.2 $^{+1.6}_{-0.9}$)~10$^{-5}$~eV$^2$ and $\tan^2\theta$~=~0.46$^{+0.04}_{-0.03}$), but
the LOW region is strongly excluded at $\Delta \chi^2 >
190$. Therefore, after the inclusion of the Borexino day--night data,
solar neutrino data alone can single out the LMA solution with very
high confidence, without the inclusion of anti-neutrino data and therefore
without invoking CPT symmetry. 

This result is an essentially new and independent way to probe the MSW-LMA 
prediction and is potentially sensitive to new physics affecting low energy electron 
neutrino interactions. As an example, we note that our day-night asymmetry measurement 
is very powerful in testing mass varying
neutrino flavor conversion scenarios. We find, for example, that our
$A_{dn}$ data excludes the set of MaVaN parameters chosen in~\cite{bib:mavan} to fit all neutrino data at more than 10$\,\sigma$.

In conclusion, we have searched for a day--night asymmetry in
the interaction rate of 862 keV $^7$Be solar neutrinos in
Borexino. The result is A$_{dn}$ = 0.001 $\pm$ 0.012 (stat) $\pm$
0.007 (syst), consistent both with zero and with the prediction of
the LMA-MSW neutrino oscillation scenario. With this result, the LOW
region of MSW parameter space is, for the first time, strongly
disfavored by solar neutrino data alone. The result constrains 
certain flavor change scenarios involving new physics.

$~~~~$This work was funded by INFN and MIUR PRIN 2007 (Italy), NSF (USA),
BMBF, DFG, and MPG (Germany), NRC Kurchatov Institute (Russia), and MNiSW (Poland). We gratefully acknowledge the generous support of the Laboratori Nazionali del Gran Sasso.

This work is dedicated to the memory of our friend Raju Raghavan, the
father of this experiment, and a great scientist.


\begin{thebibliography}{100}
%
\bibitem{bib:RadioChemical} B. T. Cleveland {\it et al.}, Astrophys. J. 496 (1998), 505;
K. Lande and P.Wildenhain, Nucl. Phys. B, Proc. Suppl. 118 (2003), 49;
W. Hampel {\it et al.} (GALLEX Coll.), Phys. Lett. B447 (1999), 127; 
J. N. Abdurashitov {\it et al.} (SAGE Coll.), Phys. Rev. Lett. 83 (1999), 4686;
 M. Altmann {\it et al.} (GNO Coll.), Phys. Lett. B 616 (2005), 174.
 %
\bibitem{bib:SK} K. S. Hirata {\it et al.} (KamiokaNDE Coll.), Phys. Rev. Lett. 63 (1989), 16; 
J. Hosaka {\it et al.}  (SuperKamiokaNDE Coll.),  Phys. Rev. D 73 (2006), 112001;
J. P. Cravens {\it et al.} (SK Coll.), Phys. Rev. D 78 (2008), 032002; 
K. Abe {\it et al.}  (SK Coll.),  Phys. Rev. D 83 (2011), 052010.

\bibitem{bib:SNO} Q. R. Ahmad {\it et al.} (SNO Coll.), Phys. Rev. Lett. 87 (2001), 071301; 
Q.~R.~Ahmad {\it et al.}  (SNO Coll.),  Phys. Rev. Lett. 89 (2002), 011301;
B. Aharmim {\it et al.} (SNO Coll.), Phys. Rev. Lett.  101 (2008), 111301; 
B. Aharmim {\it et al.} (SNO Coll.), Phys. Rev. C 81 (2010), 055504.

\bibitem{bib:Kamland} A. Gando {\it et al.} (KamLAND Coll.), Phys. Rev. D 83 (2011), 052002.

\bibitem{bib:MSW} S. P. Mikheyev and A. Yu. Smirnov, Sov. J. Nucl. Phys. 42 (1985), 913; 
L. Wolfenstein, Phys. Rev. D 17 (1978), 2369; 
K. Nakamura {\it et al.} (Particle Data Group) J. Phys. G 37 (2010), 075021.

\bibitem{bib:Regener} J. Bouchez {\it et al.}, Z. Phys. 32 (1986), 499; 
M. Cribier {\it et al.}, ÊPhys. Lett. B 182 (1986), 89;  
A.J. Baltz {\it et al.} Phys. Rev. D35 (1987); S. T. Petcov, Phys. Lett. B 434 (1998), 321. 

\bibitem{bib:SNOdn}
B.~Aharmim {\it et al.}  (SNO Coll.), Phys.\ Rev.\  C 72 (2005), 055502.

\bibitem{bib:SKdn} 
M.B. Smy {\it et al.}  (SuperKamiokaNDE Coll.) Phys. Rev. D 69 (2004), 011104.

\bibitem{bib:day-night-LMA} J. N. Bahcall {\it et al.}, JHEP 04 (2002), 007; 
A. de Gouv$\hat{e}$a {\it et al.} JHEP 03 (2001), 009.

\bibitem{bib:mavan}  P.~C.~de Holanda,  JCAP {\bf 0907} (2009), 024.


\bibitem{bib:BxBe3} G. Bellini {\it et al.} (Borexino Coll.) Phys. Rev. Lett. 107 (2011), 141302; 
arXiv:1104.1816v1 (hep-ex)

\bibitem{bib:nim1} G. Alimonti {\it et al.} (Borexino Coll.), Nucl. Instrum. Methods Phys. Res. A 600 (2009), 568.
\bibitem{bib:nim2} G. Alimonti {\it et al.} (Borexino Coll.), Nucl. Instrum. Methods Phys. Res. A 609 (2009), 58.
\bibitem{bib:muon-paper} G. Bellini {\it et al.} (Borexino Coll.) JINST 6 (2011) P05005; 
arXiv:1101.3101v2 [physics.ins-det].

\bibitem{bib:Be7Long} G. Bellini {\it et al.} (Borexino Coll.), article in preparation with details about $^7$Be data analysis, day--night search and solar neutrino oscillations analysis.

\bibitem{bib:solar-model} J.N. Bahcall, Phys. Rev. Lett. 12 (1964), 300;
C.~Pe\~na-Garay and A.~Serenelli arXiv:0811.2424 [astro-ph] (2008).
\bibitem{bib:sun-metallicity} A.~Serenelli, W.~Haxton, and C.~Pe\~na-Garay, arXiv:1104.1639 [astro-ph].
\bibitem{bib:BxB8} G. Bellini {\it et al.} (Borexino Coll.) Phys. Rev. D 82 (2010), 033006; 
arXiv:0808.2868v3 (astro-ph).

\end{thebibliography}
\end{document}